\documentclass[aps,prd,onecolumn,nofootinbib,groupedaddress,amsfonts,floatfix]{revtex4} 
\usepackage{graphicx,amsmath,amssymb,amstext}
\usepackage{amssymb,amsbsy,amsfonts,amsthm,color}

\usepackage{epsfig}
\usepackage{graphicx}
\usepackage{subfigure}

\graphicspath{{Figures/}}

\begin{document}

\title{Energy carries Information}

\author{Irfan Ilgin and I-Sheng Yang}
\affiliation{IOP and GRAPPA, Universiteit van Amsterdam, \\
Science Park 904, 1090 GL Amsterdam, Netherlands
}

\begin{abstract}
We show that for every qubit of quantum information, there is a well-defined notion of ``the amount of energy that carries it'', because it is a conserved quantity. This generalizes to larger systems and any conserved quantites: the eigenvalue spectrum of conserved charges has to be preserved while transferring quantum information. It is possible to ``apparently'' violate these conservations by losing a small fraction of information, but that must invoke a specific process which requires a large scale coherence. We discuss its implication regarding the black hole information paradox.
\end{abstract}

\maketitle

\section{Introduction and Summary}

In the recent dicussion of black hole information problem \cite{Haw76a,Mat09}, quantum information plays an increasingly important role \cite{HayPre07,BraPir09,AMPS,HH}. There is a tendency to imagine that at the level of gedanken experiments, quantum information of the real world could be manipulated as if inside a quantum computer. This impression is somewhat over-simplified and might lead to shaky conclusions. 

The quantum field theory description of the real world is not simply quantum mechanics on a high dimensional Hilbert space. Mathematically, quantum mechanics on a Hilbert space of dimension $2^N$ is about all unitary operators acting on it, namely the group $SU(2^N)$. This is similar to how we model a quantum computer with $N$ qubits. However, in the real world evolution, not all operators in $SU(2^N)$ are allowed. In fact, usually a very small subset of them can really happen. One reason is that the real world evolution conserves energy\footnote{If the Hamiltonian manifestly does not conserve energy, that means part of the dynamics is not described quantum-mechanically. One should always be able to ``enlarge'' the description to reach an energy conserving Hamiltonian.}. If the energy spectrum of the Hilbert space has no degeneracy, then the allowed unitarity operators form a much smaller subgroup $(U(1))^{2^N-1}\subset SU(2^N)$ which is totally boring. Different energy eigenstates basically do not mixed at all. When there are degenerate energy eigenstates, internal rotations within those subspaces are allowed, but it is still a subgroup much smaller than the full $SU(2^N)$.

In other words, the quantum information in the real world can only be manipulated in a very restricted way compared to processes we imagine possible in a quantum computer. In particular, ``first load the information into a quantum computer (or an auxiliatry system \cite{AMPSS}), then do whatever we want'', is an empty statement\footnote{The number of times this or similar statements appear in the recent discussion of information paradox is one of the original motivation of this paper.}. A quantum computer is part of the real world and is subjected to the same restriction. Specifically, this restriction is the following simple but surprising fact:

{\it There is a well defined notion of energy associated with every qubit of information, and it is a conserved quantity.}

In the main text, we first provide the natural definition of the energy that carries a qubit of information, which is the maximum energy difference between the two possible states. Its conservation is straightforward to prove. This means that in the evolution of the real world, a specific qubit of quantum information is always carried by a fixed amount of energy. In fact, this immediately generalizes to larger systems and all other conserved quantities. Quantum information is associated with the eigenstate spectrum of all Noether charges. 

Note that all these conservation laws impose strong constraints on how information can flow and be processed, but this fact does not imply a real difficulty in practical quantum computation. For purely computational purposes, we can encode the information in a highly degenerate system to avoid these constraints. However, if we are talking about the quantum information content within some natural system, then it comes with its own energy spectrum. If one takes the attitude that ``after loading into a quantum computer, any computation is allowed'', then it is the loading process which has to obey these constraints.

This is exactly the simplification we will adopt in this paper. We assume that gedanken experiments in the black hole information problem contain two steps: (1) loading information into a quantum computer, (2) process within the quantum computer. The information in the quatum computer is assumed to be stored in totally degenerate states, thus free of all constraints. The loading process is subjected to conservation laws, therefore not all information will be successfully loaded. We will show that in order to load a qubit of energy $E_{\rm qubit}$ and only lose a small fraction, $\epsilon\ll1$, the loading process must be coherent on a large energy scale $E_{\rm loader}\sim (E_{\rm qubit}/\epsilon)$. In other words, if there is a bound on how large an energy range is for physical processes to remain coherent, then any flow of energy larger than this bound must carry information away\footnote{Note that the inverse is not necessarily true. Information can flow without an apparent energy flow \cite{Unr12}, since some information can be stored in degenerate states and they are free to move without energy changes. Therefore, our result cannot establish a bound on how fast information flows without further specifications of the system, which agrees with the classic line of work by Pendry \cite{Pen83}.}.

If the information one would like to load belongs to the early Hawking radiation from a black hole, we will show that the loading process needs to be coherent for Planckian energy, $E_{\rm loader} = 1$. One might suspect that this is beyond the framework of QFT in a fixed background. When Planckian energy is involved in a quantum process, quantization of gravity becomes important. Any coherent process of Planckian energy migth secretly emit hard gravitons and decohere.

It was shown that without a quantum computation process, the new paradox formulated in \cite{AMPS} still fails to be observable \cite{IlgYan13}. The concern about computation time might help to resolve the paradox \cite{HH}, but it works on a time scale much longer than the evaporation time, so may not be the most relevant physics. The energetic concern might be a more direct obstacle to any future attempt to revive an observable paradox.

\section{The Energy of one qubit}
\label{sec-energy}

Consider a quantum mechanical system $A$. The way to track quantum information within this system is to imagine a reference system $R$ that is maximally entangled with $A$. For example, $R$ can be one qubit, and
\begin{equation}
|\psi\rangle_{\rm total} = \frac{1}{\sqrt{2}}\bigg(
|0\rangle_R|\psi_-\rangle_A + |1\rangle_R|\psi_+\rangle_A\bigg)~,
\label{eq-entangle}
\end{equation}
such that the system $A$ carries the one qubit of information that purifies $R$. If we select this particular basis of $R$, $0$ and $1$, then it looks like the information is carried by two orthogonal states $\psi_\pm$ in $A$. These two information-carrying states will evolve according to the dynamics of $A$, $|\psi_\pm\rangle\rightarrow U|\psi_\pm\rangle$, but such evolution conserves energy.
\begin{equation}
E\equiv \langle\psi|H|\psi\rangle_A = \langle\psi|U^\dag HU|\psi\rangle_A~.
\label{eq-conserve}
\end{equation}
This suggest that we can define the energy carrying this qubit of information as
\begin{equation}
\Delta E \equiv 
\langle\psi_-|H|\psi_-\rangle_A-\langle\psi_+|H|\psi_+\rangle_A~,
\label{eq-energyinfo}
\end{equation}
which is invariant under the dynamics of $A$. 

Note that this shall not be the final definition yet, since $\psi_\pm$ is attached to the choice of basis in $R$. A rotation in $R$ leads to a rotation in $A$.
\begin{eqnarray}
|0\rangle &\rightarrow& 
\cos\theta|0\rangle + \sin\theta|1\rangle~, \\ \nonumber
|1\rangle &\rightarrow& 
\sin\theta|0\rangle - \cos\theta|1\rangle~, \\
|\psi_+\rangle &\rightarrow& 
\cos\theta|\psi_+\rangle + \sin\theta|\psi_-\rangle~, \\ \nonumber
|\psi_-\rangle &\rightarrow& 
\sin\theta|\psi_+\rangle - \cos\theta|\psi_-\rangle~. 
\end{eqnarray}
The energy defined in Eq.~(\ref{eq-energyinfo}) will be a sinusoidal function of $2\theta$. The natural definition of the energy that carries this qubit of information should be the amplitude of this function. In other words, 
\begin{equation}
E_{\rm qubit} = \Delta E(\theta_{\rm max})~,
\label{eq-energymax}
\end{equation}
where $\theta_{\rm max}$ is the choice of basis in $R$ that maximizes Eq.~(\ref{eq-energyinfo}).

The physical meaning of this energy is quite simple. The operational definition of the maximal entanglement between $A$ and $R$ is that for every basis in $R$, there will be a corresponding two-states projection measurement in $A$ such that their results are exactly correlated. In other words, if all of those projections are doable in $A$, then we can use them to predict the outcome of any measurement in $R$. That includes a measurement of $R$ in this particular basis $\theta_{\rm max}$. So, the projection in $A$ must allow two possible outcomes with energy difference given by Eq.~(\ref{eq-energymax}). In our description, $A$ is a large system that conserves total energy (maybe the entire world). One often wants to see if a subsystem $A'\in A$ contains the entire information that purifies $R$. For that purpose, we know that the subsystem $A'$ cannot contain such quantum information if it does not have two states with energy difference given by Eq.~(\ref{eq-energymax}).

In fact, there can be other conserved Noether charges in the Hamiltonian, such as momentum and various quantum numbers. It is straightforward to generalize our definition to all of them. A given qubit of information is carried by all these Noether charges it started with. Note that the choices of basis in $R$ to maximize the differences in various Noether charges will be different. 

\section{Loading a quantum computer}
\label{sec-load}

Note that our argument in the previous section was for the ``entire'' information. For practical purposes, one should consider the possibility to preserve ``most'' of the information while letting go some or all of its energy. Such a process will allow us to approximately load a qubit of information with energy  $E_{\rm qubit}>0$ into a ``computational'' qubit with zero energy. After that, the computation process is no longer constrained by energy conservation. That is in-principle possible, but a special loading process is required.

This loading process still obeys conservation of energy, which implies that it must be something like
\begin{equation}
|\psi_\pm\rangle_A |\phi_0\rangle_{\rm comp} 
|\Phi_0\rangle_{\rm loader} \rightarrow 
|\psi_0\rangle_A
|\phi_\pm\rangle_{\rm comp} |\Phi_\pm\rangle_{\rm loader}~.
\label{eq-load}
\end{equation}
When the qubit of information $\psi_\pm$ in system $A$ is loaded into the degenerate quantum computer states $\phi_\pm$, the conserved energy difference must be carried away by the loader states $\Phi_\pm$. In addition, despite having different energies, the two loader states $\Phi_\pm$ cannot be too different. If they are distinquishable from each other, the state of the loader will be somewhat entangled with the state of the quantum computer. Such entanglement undermines the information transfer from system $A$ to the quantum computer. 

In order to quantify how much information is successfully transfered, let us assume that the qubit to load from $A$ was in a pure state, 
\begin{equation}
|\psi\rangle_A=(|\psi_+\rangle+|\psi_-\rangle)/\sqrt{2}~.
\end{equation} 
This will be loaded to
\begin{equation}
\frac{1}{\sqrt{2}}\left(|\phi_+\rangle_{\rm comp} |\Phi_+\rangle_{\rm loader}+|\phi_-\rangle_{\rm comp} |\Phi_-\rangle_{\rm loader}\right)~.
\end{equation}
Tracing over the loader system, we get the density matrix of the quantum computer,
\begin{equation}
\rho_{\rm comp}=
\left( \begin{array}{cc}
1 & \langle\Phi_+|\Phi_-\rangle  \\
\langle\Phi_-|\Phi_+\rangle & 1  \end{array} \right)~.
\end{equation}
In order for the information to be almost fully loaded, this should be close to a pure state. That requires
\begin{equation}
|\langle\Phi_-|\Phi_+\rangle| = 1-\epsilon~,
\end{equation}
where $\epsilon\ll1$ is small number which parametrizes ``how much information is lost'' during the loading process. A small $\epsilon$ implies that the two loader states $\Phi_\pm$ are almost indistinguishable from each other, thus not very entangled with the quantum computer. 

We will demonstrate that a small $\epsilon$ requires the following three conditions:
\begin{itemize}
\item The energy spectrum of the loader system is dense on the scale of $E_{\rm qubit}$.
\item The loader states $\Phi$ have large uncertainties in energy, $\Delta E_{\rm loader} \gg E_{\rm qubit}$.
\item The loading process is coherent for many microstates spanning energy range $\Delta E_{\rm loader}$.
\end{itemize}
This is most easily visualized by the following construction\footnote{Such construction is inspired by how Aharonov and Susskind circumvented the superselection rule \cite{AhaSus67}.}. For convenience, we set $\psi_0=\psi_-$, $\phi_0=\phi_-$, $\Phi_0=\Phi_-$ in Eq.~(\ref{eq-load}). This means that an input state of $|\psi_-\rangle_A$ has no effect on the loader.
\begin{eqnarray}
|\Phi_0\rangle = |\Phi_-\rangle = \sum_{n=1}^N N^{-1/2}|E_n\rangle~.
\end{eqnarray}
$|E_n\rangle$ are energy eigenstates of the loader, and a dense spectrum allows us to pick $E_n-E_{n-1} = E_{\rm qubit}$ for all $n$. This arrangement is necessary because when the input is $|\psi_+\rangle_A$, the loader state can change to the following form: 
\begin{eqnarray}
|\Phi_+\rangle = \sum_{n=2}^{N+1} N^{-1/2}e^{i\theta_n}|E_n\rangle~.
\end{eqnarray}
This is demanded by energy conservation for a unitary transformation. Since system $A$ loses energy $E_{\rm qubit}$ and none of which goes into the quantum computer, every energy eigenstate in the loader system needs to pick up such energy and shift to the next energy eigenstate.

The other two requirements become obvious as we compute
\begin{equation}
|\langle\Phi_-|\Phi_+\rangle| = 
\left|\sum_{n=2}^{N} e^{i\theta_n}N^{-1}\right|~.
\end{equation}
If the interaction with the loader is not coherent for these many microstates, the random phases $e^{i\theta_n}$ will cancel each other \cite{tHo12}. That means $\langle\Phi_-|\Phi_+\rangle\sim0$. In such case the loader is maximally entangled with the quantum computer, and the information from $A$ fails to be loaded. The third condition we stated earlier was to prevent this from happening.

After those phases $\theta_n$ are aligned, a large $N$ implies that $\epsilon\sim N^{-1}$ is small. Since the loader state needs to be a superposition of many different energy eigenstates, it has a large uncertainty in energy as we stated in the second condition.

As a quick summary: In order to lose a small fraction of information $\epsilon$ during the loading process, we need the loader system to be dense and coherent on a large energy range, 
\begin{equation}
\Delta E_{\rm loader} \sim NE_{\rm qubit}
\sim E_{\rm qubit}/\epsilon\gg E_{\rm qubit}~.
\end{equation}

Note that it is difficult for a system to have a large uncertainty but still interacts coherently. One usually creates a large uncertainty in some variable by measuring its conjugate variable very accurately. Such uncertainty comes with a large entanglement and cannot maintain the coherence of the loading process. For example, if the loader is in a mixed state,
\begin{equation}
\rho_{\rm loader} = \sum_{n=1}^N N^{-1}
|E_n\rangle\langle E_n|~,
\end{equation}
then it does have a large uncertainty. However this loader state will make $\rho_{\rm comp}$ diagonal, which means that all information is lost during the loading process.

This means that designing such a good loader is not only technically, but also thoeretically challenging. It is fundamentally different from other auxilliary systems appear in the discussion of quantum computation, such as the ancilla for error correction or the waste during computation. Those auxilliary systems can be reset into its initial form and keep functioning. A loader, on the other hand, is hard to reset since interacting with another system most likely undermines the coherent condition and renders it useless.

\section{Hawking radiation}
\label{sec-OBH}

In the context of black hole information problem, the information one would like to load into a quantum computer belongs to the ``early Hawking radiation''. For a black hole of mass $M$, the early Hawking radiation is roughly $M^2$ qubits of energy $E_{\rm qubit}\sim M^{-1}$, where the Planck unit is set to one. Loading one qubit is still easy, which requires the loader to be coherent for $\Delta E_{\rm loader}\gg M^{-1}$. However, without resetting the loader, it eventually needs to load all $M^2$ qubits, which requires coherence for some macroscopic energy $\Delta E_{\rm loader}\gg M^2M^{-1} = M$. 

This estimation probably exaggerates the difficulty of loading. The energy difference $M^2M^{-1}$ can only occur between atypical states in the system of early radiation, for example between the state of almost $M^2$ unoccupied qubits and almost $M^2$ occupied qubits. To be more fair, the energy that stores these $M^2$ qubits of energy should only be associated with the energy uncertainty of typical states, $E_{M^2~{\rm qubits}}\sim \sqrt{M^2}E_{\rm qubit}\sim1$, which happens to be the Planck scale\footnote{Note that Planckian energy is the natural fluctuation of total mass for a black hole in the canonical ensemble. Even if we use a micro-canonical ensemble to limit the total mass of the original black hole, fluctuation will grow during the evaporation process, reach Planckian when it is half-evaporated, and stay at that value to almost the end.}.

The exact physical meaning of this fact is open for discussion. In the context of black hole quantum information \cite{HayPre07}, the important question is whether this Planckian qubit of information should be considered as part of the low energy physics? It would seem like in order to describe a physical process which remains coherent in such energy scale, quantization of gravity is inevitably important. Note that we are objecting to the fact that the early Hawking radiation itself has a Planckian energy uncertainty. It is natural for a large physical system to accumulate a large uncertainty in any observable. We are arguing that any attempt to extract a specific fraction of information from the early radiation (known as the distillation process \cite{Sus13}) is facing an obstacle. That is because any such process is effectively ``first load into a quantum computer, then perform quantum computation''. In this case it must invoke a physical process which is coherent across Planckian energy scale. 

\section{Discussion}

The core argument of the black hole information paradox is quantum cloning (or equivalently, violating the monogamy of entanglement). In order to challenge complementarity, the duplicated quantum information must be demonstrated to reside within one causal patch \cite{SusTho93,HayPre07,SekSus08}. As we recently shown in \cite{IlgYan13}, if the information that purifies a late Hawking quantum remains distributed across the entire early Hawking radiation, then no cloning exists within any causal patch. Conceptually speaking, the entire early radiation is too bulky, and it is impossible to fit them into an infalling causal patch.

The remaining hope to establish a paradox is the quantum computation process proposed in \cite{AMPS}, and it is still an open issue. There are various different concerns about its validity, such as the time it requires \cite{HH} and the back-reaction it causes \cite{HuiYan13}. The argument using an auxiliary system (such as the setup in \cite{AMPSS}) seems to circumvent these concerns, but it brings the question to another level which is no longer the validaty of semi-classical, four-dimensional physics.

In this paper, we study another possible issue for quantum computation in this context which is realted to energy. In both \cite{IlgYan13} and here, we have followed the formulation of the paradox \cite{AMPS} and assumed that the black hole is entangled with its own radiation. However, the only necessary input is actually energetic properties: that a black hole of mass $M$ is entangled with $N=M^2$ qubits of energy $M^{-1}$. All of our arguments will also work if the black hole is entangled with any system with higher total energy, which obviously can happen if Hawking radiation interacts with a larger system. It would not have worked if a black hole is entangled with a less energetic system: the same number of particles at a much lower temperature. Fortunately, the second law of thermodynamics seems to forbid such situation. If we follow the recipe in \cite{OppUnr14} to make a ``pre-entangled'' black hole, it can only be entangled with hotter gas but not colder.

The association between energy and information is the core idea behind the second law, and a careful analysis in both provides solutions to classical information paradoxes such as Maxwell's Demon \cite{Llo96,MarNor09}. In the recent discussions of black hole information paradox, energetic concerns have been somewhat ignored. We hope that by pointing out the definite connection between energy and quantum information, we can get one step closer to the resolution of the paradox.

\section{acknowledgments}

We thank Jacob Bekenstein, Raphael Bousso, Ben Freivogel, Daniel Harlow, Don Page and Erik Verlinde for discussions. We are particular grateful that Douglas Stanford pointed out a technical mistake in the previous version. We also thank the hospitality of the Weizmann Institute during the workshop ``Black Hole and Quantum Information''. This work is supported in part by the Foundation for Fundamental Research on Matter (FOM), which is part of the Netherlands Organization for Scientific Research (NWO).

\bibliography{all}

\end{document}